\documentclass[12pt]{article}
\usepackage[dvipdfmx]{graphicx}
\begin{document}
\title{Classicality in Quantum Mechanics: \\ model for pointer states and decoherence}
\author{Kentaro Urasaki\\
Tokyo, Japan, urasaki@ynu.ac.jp}
\date{}%

\maketitle

\abstract{
We have studied the emergence of classical states in the perturbative interaction model. 
The states which interact with many other degrees of freedom, 
such as the center of mass of a macro-object, play important role.
Although the random phase mechanism is effective as same as Zurek's strong correlation model\cite{Zurek1982}, there are enormous states, each of which independently developes due to 
the orthogonality of the environmental states.
In these privileged states, the subsystem picture with the separability is stable.
}


\section{Introduction: instability of the pointer states}

\subsection{Classicality from quantum unitarity}

One of the remarkable properties of classical objects
is that a physical quantity always has a cetain value. 
Such the classical quantities, however, are rare, while 
the almost all of other degrees of freedom are under the quantum unitary evolution. 

Recent studies suggest that the part of the classical nature of macro-objects may originate 
from the unitary evolution of quantum mechanics.
It seems also reasonable since
quantum mechanics is ahead of classical mechanics
from the viewpoint of cosmology.
In addition, it is very interesting if the classical states emerges from
the entanglement of microscopic quantum degrees of freedom.
To clarify the physical condition for this mechanism and 
how classical degrees of freedom interact with the other quantum 
degrees of freedom
will be important step to understand the emergence of the order in the universe.

\vspace{1cm}

Among these studies, the decoherence theory\cite{Joos1985}\cite{Zurek1981} insists that the interaction between a system and the environment is relevant to some parts of the classicality. 
The classicality in this context means that the quantum coherent terms are effectively lost 
in the physical quantities 
when it is expressed by the privileged set of states. 
This scenario is considered to explain, for example, 
that the observed center of mass (COM) of macro-objects are always localized in space.

\subsection{Decoherence scenario}

For simplicity, we treat the two orthogonal states $|\phi_1\rangle$ and $|\phi_2\rangle$ of a macroscopic system $\phi$ interacting with its environment $\varepsilon$.
Let us start from the following initial state of the total system:
\begin{eqnarray}\label{vNform-1}
|\Phi(0)\rangle=(c_1|\phi_1(0)\rangle+c_2|\phi_2(0)\rangle)|\varepsilon(0)\rangle.
\end{eqnarray}
If we can assume an appropriate interaction, 
it evolves into 
\begin{eqnarray}\label{vNform-2}
|\Phi(t)\rangle=c_1|\phi_1(t)\rangle|\varepsilon_{\phi_1}(t)\rangle+c_2|\phi_2(t)\rangle
|\varepsilon_{\phi_2}(t)\rangle,
\end{eqnarray}
where two states, $|\phi_1(t)\rangle, |\phi_2(t)\rangle$, are stable against the interaction (Hamiltonian) 
and called as the pointer states which play essential role in the decoherence scenario.

For the operator $\hat{Q}$ acting only on the subsystem $\phi$, 
its expectation value is expressed as, 
\begin{eqnarray}
\langle Q \rangle=\!\!\!\!&&|c_1|^2\langle\phi_1(t)|\hat{Q}|\phi_1(t)\rangle+|c_2|^2\langle\phi_2(t)|\hat{Q}|\phi_2(t)\rangle\\
&&+c^\ast_1 c_2\langle\phi_1(t)|\hat{Q}|\phi_2(t)\rangle\langle\varepsilon_{\phi_1}(t)|\varepsilon_{\phi_2}(t)\rangle
+c^\ast_2 c_1\langle\phi_2(t)|\hat{Q}|\phi_1(t)\rangle\langle\varepsilon_{\phi_2}(t)|\varepsilon_{\phi_1}(t)\rangle.
\end{eqnarray}
If the state of the environment develop into the corresponding orthogonal state
after the interaction, we can say, 
\begin{eqnarray}
\langle\varepsilon_{\phi_1}(t)|\varepsilon_{\phi_2}(t)\rangle\sim 0.
\end{eqnarray}
In this case, the coherent terms in the above equation can be neglected. 
This process is called decoherence.
Since in this case the above equation is similar to that of statistical mixture of the events, 
we can also say the {\it approximate} mixture is obtained. 
We also find out the same result  
introducing the reduced density operator for the macroscopic system as,
\begin{eqnarray}
\rho_\phi:={\rm Tr}_\varepsilon\rho_\Phi,
\end{eqnarray}
where the density matrix for mixed states in the privileged Schmidt {\it basis}, $|\phi_1\rangle, |\phi_2\rangle$, appears. 
These non-obvious results from the unitary evolution explain some part of the classicality.

The expansion of the range of the prediction by the quantum unitarity is relevant to  
the interpretation.
The total state, however, is still in the superposition of these states.
Therefore the gap still exists between the quantum unitary evolution and our daily experience in which the single result stochastically emerges. 
[Schlosshauer]

As surveyed in the next subsection, in this scenario, 
the interaction Hamiltonian should not only be approximately diagonal in $|\phi_1(t)\rangle, |\phi_2(t)\rangle$ but also
commute with the system's self-Hamiltonian 
since the non-commutativity destroys the separability of the subsystems $\phi$ and $\varepsilon$. At present, the above results originate from the reduction of the environmental degrees of freedom. 
We see why the decoherence scenario hardly work in the state vector description below.

\subsection{Fragility of subsystems}

The model for the quantum measurement by von Neumann\cite{Neumann1932} and the model for the quantum decoherence by Zurek\cite{Zurek1982} are important to refer 
the quantum mechanical approach to macroscopic systems.
Although these leading studies clarify the importance of the branch structure, 
eq. (\ref{vNform-2}),
this structure strongly depends on the assumption of the commutativity 
between the system and the interaction Hamiltonian.
Above two models satisfy this condition (namely $[\hat{h}_0, \hat{h}_I]\simeq 0$)
by assuming the instant interaction or the strong correlation limit.

To understand this problem from the subsystem picture of the system, 
let us use the perturbative aspect. 
Leaving the aspect of  ``a system and its environment'' aside,
we fully expand the one branch by the eigenstates of the interaction Hamiltonian:
\begin{eqnarray}
|\phi_1(0)\rangle|\varepsilon(0)\rangle&&=
|\phi_1\rangle(c_{11}|\varepsilon_1\rangle+c_{12}|\varepsilon_2\rangle+\cdots+c_{1N}|\varepsilon_N\rangle)\\
&&=c_{11}|\phi_1\rangle|\varepsilon_1\rangle+\cdots+c_{1N}|\phi_1\rangle|\varepsilon_N\rangle,\\
|\phi_2(0)\rangle|\varepsilon(0)\rangle&&=
|\phi_2\rangle(c_{21}|\varepsilon_1\rangle+c_{22}|\varepsilon_2\rangle+\cdots+c_{2N}|\varepsilon_N\rangle)\\
&&=c_{21}|\phi_2\rangle|\varepsilon_1\rangle+\cdots+c_{2N}|\phi_2\rangle|\varepsilon_N\rangle.
\end{eqnarray}
In the case that the self-Hamiltonian $\hat{h}_0$ is the main term and 
the interaction Hamiltonian $\hat{h}_I$ is perturbative, 
we should firstly consider the transition between $\phi_1$ and $\phi_2$.
Namely, due to the self-evolution by $\hat{h}_0$, 
the leading time dependence of $|\phi_1\rangle$ is $|\phi_1(t)\rangle=\alpha_{11}(t)|\phi_1\rangle+\alpha_{12}(t)|\phi_2\rangle$ (in the Schmidt basis).
On the other hand, $|\phi_2(t)\rangle=\alpha_{21}(t)|\phi_1\rangle+\alpha_{22}(t)|\phi_2\rangle$. 

Since this necessarily causes the interaction between two branches,
the each branch is not even approximately closed:
the off-diagonal element is the same order of the diagonal element, where the latter is expected to cause the random phase effect:
\begin{eqnarray}
\langle\phi_1(t)|\langle\varepsilon_1(t)|\hat{h_I}|\phi_2(t)\rangle|\varepsilon_1(t)\rangle\sim
\langle\phi_1(t)|\langle\varepsilon_1(t)|\hat{h_I}|\phi_1(t)\rangle|\varepsilon_1(t)\rangle.
\end{eqnarray}
This means, as in the textbook-quantum-mechanics, $|\phi_1(t)\rangle|\varepsilon_1(t)\rangle$ and 
$|\phi_2(t)\rangle|\varepsilon_1(t)\rangle$ are not independent states in the time evolution.
At the sametime, the continuous transition destroys the separability of the subsystems, 
$\phi$ and $\varepsilon$.
This is the general consequence brought on by the non-commutativity, $[\hat{h}_0, \hat{h}_I]\ne 0,$
which is relevant to the fragility of the subsystem picture of interacting quantum systems.
\footnote{Although the discussions using 
the reduced density matrix do not directly need such the branch structure,
the absence of the approximately closed subsystems probably makes 
the decohrence scenario meaningless.}

Therefore, only in the limit of the strong interaction, the subsystem picture and the random phase mechanism seem to be compatible with the branch of the eigenstates of the interaction Hamiltonian.
Zurek's model\cite{Zurek1982} gives important suggestion at this point; 
the random phase mechanism for the eigenstate of the interaction Hamiltonian 
makes the pointer states emerge under the strong interaction. 
Below we, however, will try to re-find the branch structure consisting of the subsystems of a quantum state under the perturbative interaction.

\newpage

\section{Pointer states for the center of mass}

\subsection{Model for center-of-mass: self-evolution}

Such as a COM under non-uniform potential, 
some types of degrees of freedom weekly interact 
with many other inner degrees of freedom.
The Fig. 1 represents the model to investigate in the present paper.
The system $|\phi\rangle$ interacts with every $|\sigma\rangle_l$, $(1\le l\le M)$,
while $|\sigma\rangle_l$'s have no interaction with each other.

\vspace{1cm}
\underline{\bf Model}\\

For easy treatment, we assume $\phi$ and each $l$ consist of two states system.
The environmental state is represented as $|\varepsilon\rangle=|\ \ \rangle_1|\ \ \rangle_2\cdots|\ \ \rangle_M$,
and can be expressed by $N=2^M$ basis.

\begin{figure}
\begin{center}
\includegraphics[width=7cm]{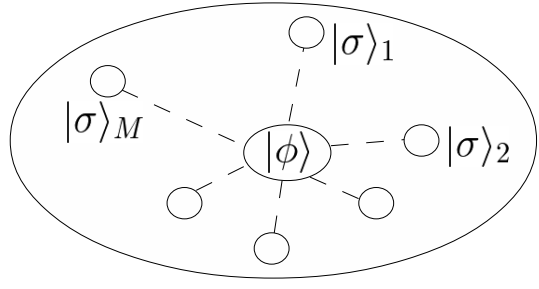}
\caption{The system and the environment (image).}
\end{center}
\end{figure}

To study the case that the interaction Hamiltonian and the self-Hamiltonian are not commutative,
in the eigenstates of the interaction Hamiltonian, $|\phi_1\rangle$ and $|\uparrow\rangle_l$, 
we set the self-Hamiltonian as,  
\begin{eqnarray}
\hat{h}_{\phi}=E\ (|E_{+}\rangle\langle E_{+}|-|E_{-}\rangle\langle E_{-}|),\\
\hat{h}_l=\omega_l(|+\rangle\langle+|_l-|-\rangle\langle-|_l).
\end{eqnarray}
\begin{eqnarray}
|E_\pm\rangle=\frac{1}{\sqrt{2}}(|\phi_1\rangle\pm|\phi_2\rangle),\\
|\pm\rangle_l=\frac{1}{\sqrt{2}}(|\uparrow\rangle_l\pm|\downarrow\rangle_l)
\end{eqnarray}
We define the self-evolution of these states:
\begin{eqnarray}
|\phi_1(t)\rangle:=&&e^{-i\hat{h}_\phi t}|\phi_1\rangle\\
=&&\frac{e^{-iE t}}{\sqrt{2}}|E_+\rangle+\frac{e^{iE t}}{\sqrt{2}}|E_-\rangle\\
=&&\cos E t |\phi_1\rangle-i\sin E t |\phi_2\rangle,\\
|\uparrow(t)\rangle_l:=&&e^{-i\hat{h}_l t}|\uparrow\rangle_l\\
=&&\cos \omega_l t |\uparrow\rangle_l-i\sin \omega_l t |\downarrow\rangle_l,
\end{eqnarray}
and so on.

The matrix element are diagonal in $|\phi_1\rangle, |\phi_2\rangle, |\uparrow\rangle_l$ 
and $|\downarrow\rangle_l$,
\begin{eqnarray}
\hat{h}_{\phi, l}=\sum_{i=1,2}\sum_{\sigma=\uparrow, \downarrow}v_{i, \sigma}^l|\phi_i\rangle\langle\phi_i|\otimes|\sigma\rangle\langle\sigma|_l,\\
\end{eqnarray}

The total Hamiltonian is, 
\begin{eqnarray}
\hat{h}=\hat{h}_0+\hat{h}_I.
\end{eqnarray}
We here define the self-Hamiltonian as,
\begin{eqnarray}
\hat{h}_0:=\hat{h}_{\phi}\otimes\hat{1}_\varepsilon+\hat{1}_\phi\otimes\hat{h}_\varepsilon,
\end{eqnarray}
where, 
\begin{eqnarray}
\hat{h}_\varepsilon:=\sum_l\hat{h}_{l}\ \prod_{l^\prime\neq l}\otimes\hat{1}_{l^\prime}.
\end{eqnarray}
On the other hand, the total interaction is defined as, 
\begin{eqnarray}
\hat{h}_I:=\sum_l\hat{h}_{\phi, l}\ \prod_{l^\prime\neq l}\otimes\hat{1}_{l^\prime}.
\end{eqnarray}
The case $E=\omega_l=0$ can be identified to Zurek's model.

\subsection{Diagonal approximation for environmental states}

In Zurek's model, the scenario that the oscillation in the phase of the state vectors 
selects out the classical states makes use of the random phase mechanism due to 
the interaction energy between the system and its environment. 
In this mechanism, it is important that the phase shift is prior to 
the transition by the interaction. 
Namely, the matrix element of the interaction Hamiltonian should be diagonal. 
In the case that the number of the environmental states is sufficiently large, 
we can expect that this condition is realized due to the orthogonality of it.

Resultingly, the state of the total system $|\phi\rangle\otimes|\varepsilon\rangle$ is localized on the eigenstate of the environment $|\varepsilon_\nu(t)\rangle$, and 
the random phase mechanism can work.

\begin{figure}
\begin{center}
\includegraphics[width=7cm]{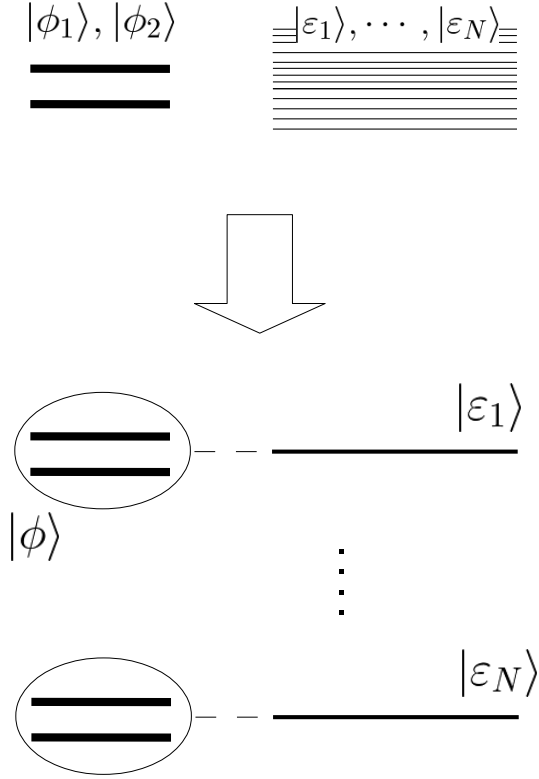}
\caption{The separation to the independently time-evoluting states (image).}
\end{center}
\end{figure}

\vspace{1cm}
\underline{\bf Calculation}\\

The initial states of the environment are labeled as,
\begin{eqnarray}
|\varepsilon_1(0)\rangle:=
|\uparrow\rangle_1 |\uparrow\rangle_2 |\uparrow\rangle_3 \cdots  |\uparrow\rangle_{M-1}|\uparrow\rangle_M, \\
|\varepsilon_2(0)\rangle:=
|\downarrow\rangle_1 |\uparrow\rangle_2 |\uparrow\rangle_3 \cdots  |\uparrow\rangle_{M-1}|\uparrow\rangle_M, \\
\cdots,\\
|\varepsilon_N(0)\rangle:=
|\downarrow\rangle_1 |\downarrow\rangle_2 |\downarrow\rangle_3 \cdots  |\downarrow\rangle_{M-1}|\downarrow\rangle_M.
\end{eqnarray}
The self-evolution is, 
\begin{eqnarray}
|\varepsilon_1(t)\rangle:=
e^{-i\hat{h}_\varepsilon t}|\uparrow\rangle_1 |\uparrow\rangle_2 \cdots |\uparrow\rangle_M
=\prod_{l=1}^{M} (\cos \omega_l t|\uparrow\rangle_l-i\sin \omega_l t|\downarrow\rangle_l),
\end{eqnarray}
and so on.
We advance toward the approximation. 
Substituting the expansion by non-perturbative states,
\begin{eqnarray}\label{expansion}
|\Phi(t)\rangle=\sum_{\nu^\prime=1}^N\sum_{j=1, 2}c_{j\nu^\prime}(t)|\phi_j(t)\rangle|\varepsilon_{\nu^\prime}(t)\rangle,
\end{eqnarray}
into the exact Schr\"odinger equation $[i\hbar\partial_t-\hat{h}]|\Phi(t)\rangle=0,$ 
\begin{eqnarray}\label{expansion-Seq}
\sum_{\nu^\prime=1}^N\sum_{j=1,2}i\hbar\ \dot{c}_{j\nu^\prime}(t)|\phi_j(t)\rangle|\varepsilon_{\nu^\prime}(t)\rangle=\hat{h}_{I}\sum_{\nu^\prime=1}^N\sum_{j=1, 2}c_{j\nu^\prime}(t)|\phi_j(t)\rangle|\varepsilon_{\nu^\prime}(t)\rangle
\end{eqnarray}
The time dependence, 
$|\phi_j(t)\rangle$ and $|\varepsilon_{\nu^\prime}(t)\rangle$, indicate the self-evolution.

\vspace{1cm}

The self-evolution does not change the orthogonality relation, 
\begin{eqnarray}
\langle\sigma(t)|\sigma^\prime(t)\rangle_{l^\prime}=\delta^{(l^\prime)}_{\sigma\sigma^\prime}, \\
\langle\varepsilon_{\nu}(t)|\varepsilon_{\nu^\prime}(t)\rangle=\delta_{\nu\nu^\prime},
\end{eqnarray}
while the matrix elements fluctuate around the initial value as, 
\begin{eqnarray}
\langle\varepsilon_\nu(t)|\hat{h}_I|\varepsilon_{\nu^\prime}(t)\rangle=
\sum_{l=1}^M\langle\sigma(t)|_l\ \hat{h}_{\phi, l}|\sigma^\prime(t)\rangle_l
\prod_{l^\prime\ne l}\delta^{(l^\prime)}_{\sigma\sigma^\prime}.
\end{eqnarray}
This, however, is not zero only when the spins of $|\varepsilon_\nu(t)\rangle$ is different 
from those of $|\varepsilon_{\nu^\prime}(t)\rangle$ just on one site.

Let us evaluate this fluctuation by the self-evolution acting 
$\langle\varepsilon_1(t)|$ from the left side of eq. (\ref{expansion-Seq}). 
We obtain the equation for the coefficients, $c_{1N}(t)$ and $c_{2N}(t)$, 
\begin{eqnarray}\label{expansion-1}
\sum_{j=1,2}i\hbar\ \dot{c}_{j1}(t)|\phi_j(t)\rangle
=\sum_{\nu^\prime=1}^N\sum_{j=1, 2}c_{j\nu^\prime}(t)|\phi_j(t)\rangle
\langle\varepsilon_1(t)|\hat{h}_{I}|\varepsilon_{\nu^\prime}(t)\rangle.
\end{eqnarray}
Of course, at $t=0,$ the matrix element, 
\begin{eqnarray}
\langle\varepsilon_1|\hat{h}_I|\varepsilon_{\nu^\prime}\rangle=\delta_{1, \nu^\prime}\sum_{l=1}^M
\hat{v}_{\uparrow\uparrow}^l
\end{eqnarray}
is diagonal.
On the other hand, at $t\ne 0$,
$\langle\varepsilon_1(t)|\hat{h}_I|\varepsilon_{\nu^\prime}(t)\rangle\ne 0$
for $1\le \nu^\prime\le M+1$.

We can easily calculate the self-evolution of the diagonal element ($\nu^\prime=1$) as, 
\begin{eqnarray}
\langle\varepsilon_1(t)|\hat{h}_I|\varepsilon_1(t)\rangle
&&=
\sum_{l=1}^M\langle\uparrow(t)|\hat{h}_{\phi, l}|\uparrow(t)\rangle_l\\
&&=
\sum_{l=1}^M\hat{v}_{\uparrow\uparrow}^l\cos^2\omega_l t
+\hat{v}_{\downarrow\downarrow}^l\sin^2\omega_l t.
\end{eqnarray}
For the off-diagonal elements ($2\le \nu^\prime\le M+1$),
\begin{eqnarray}
\sum_{\nu^\prime=2}^Nc_{j\nu^\prime}(t)\langle\varepsilon_{1}(t)|\hat{h}_I|\varepsilon_{\nu^\prime}(t)\rangle
&&=\sum_{l=1}^Mc_{jl+1}(t)
\langle\uparrow(t)|\hat{h}_{\phi, l}|\downarrow(t)\rangle_l\\
&&=\sum_{l=1}^Mc_{jl+1}(t)(\hat{v}_{\uparrow\uparrow}^l+\hat{v}_{\downarrow\downarrow}^l)
i\sin\omega_lt\cos\omega_lt.
\end{eqnarray}

At $t=0$, the diagonal element is order of $O(M)$, 
while the off-diagonal elements are exactly zero.
Here, let us assume that the phase ${\omega_l}t$ are distributing randomly around zero at $t\ne 0$.
Then the fluctuation of the on- and off-diagonal elements due to the self-evolution is order of $O(\sqrt{M})$ for sufficiently large $M.$
{\bf Therefore we can neglect the off-diagonal elements compared with the diagonal one 
even taking into account of the self-evolution.}

\vspace{1cm}

Next, we evaluate the interaction effect on $\phi,$ 
There are two types of the effect, where 
one is the transition $\phi_1\rightleftharpoons\phi_2$ and the other 
is the common phase shift. 
We here take into account only the latter effect by setting  
$c_{i\nu}(t)=\alpha_\nu(t)c_{i\nu}, $ where $|c_{1\nu}|^2+|c_{2\nu}|^2=1.$
For $\alpha_1(t),$ eq. (\ref{expansion-1}) leads, 
\begin{eqnarray}
i\hbar\ \dot{\alpha_1}(t)&&=\alpha_1(t)\lambda_1(t)-
\sum_{l=1}^{M}\alpha_{l+1}(t) \lambda_{l+1, 1}(t)\
i\sin\omega_{l} t\ \cos\omega_{l} t\\
&&\simeq \alpha_1(t)\lambda_1(t),
\end{eqnarray}
where, as mentioned above, the off-diagonal terms can be neglected for sufficiently large $M$.
This diagonal term,
\begin{eqnarray}\label{lambda}
\lambda_\nu(t):=\langle\varepsilon_\nu(t)|(c^\ast_{1\nu}\langle\phi_1(t)|+c^\ast_{2\nu}\langle\phi_2(t)|)\hat{h}_I
(c_{1\nu}|\phi_1(t)\rangle+c_{2\nu}|\phi_2(t)\rangle)|\varepsilon_\nu(t)\rangle,
\end{eqnarray}
is just the interaction energy.

\vspace{1cm}

We can rewrite the state vectors to stress the unit states which are stable against the time-evolution as, 
\begin{eqnarray}
|\nu(t)\rangle:=(c_{1\nu}|\phi_1(t)\rangle+c_{2\nu}|\phi_2 (t)\rangle)|\varepsilon_\nu(t)\rangle, 
\end{eqnarray}
and the above equation becomes, 
\begin{eqnarray}
i\hbar\dot{\alpha}_\nu(t) |\nu(t)\rangle=\hat{h}_I |\nu(t)\rangle.
\end{eqnarray}
The time dependence of $|\nu(t)\rangle$ represents the self-evolution and $\langle\nu(t)|\nu^\prime(t)\rangle=\delta_{\nu\nu^\prime}.$
Then, the perturbation on the phase shift is, 
\begin{eqnarray}
i\hbar\partial_t\alpha_\nu(t)  \simeq \alpha_\nu(t)\lambda_\nu(t),
\end{eqnarray}
where, 
\begin{eqnarray}
\lambda_\nu(t):=\langle \nu(t)|\hat{h}_I|\nu(t)\rangle.
\end{eqnarray}
We can very easily integrate this and obtain
$\displaystyle \alpha_\nu(t)=\alpha_{\nu}\exp\left[-i\int \langle \nu(t)|\hat{h}_I|\nu(t)\rangle dt/\hbar\right].$
After all, the diagonal approximation leads, 
\begin{eqnarray}
|\Phi(t)\rangle\simeq\sum_{\nu=1}^N\alpha_\nu|\nu(t)\rangle e^{-i\Lambda_\nu(t)/\hbar}, 
\end{eqnarray}
where $\displaystyle\Lambda_\nu(t):=\int\langle\nu(t)|\hat{h}_I|\nu(t)\rangle dt$
represents the time integral of the interaction energy. 

\vspace{1cm}
In short, the transition is suppressed by the diagonality of the environment:
The entangled state $|\nu(t)\rangle$ is the minimum unit 
in which the separability of the subsystems is kept, where
a kind of localization on $\nu$ enables
the multiple scattering effect on the phase of the state vectors.

\subsection{Random phase mechanism}
We assume that 
$|\phi_\nu(t)\rangle=c_{1\nu}|\phi_1(t)\rangle+c_{2\nu}|\phi_2(t)\rangle$
represents various superposition of states depending on $\nu$.
Even for the very weak interaction, $\Lambda_\nu(t)$ is the macroscopic quantity ($O(M)$) and 
occurs very frequent sign inversion on the states.
Therefore only the states giving the extreme values to $\Lambda_\nu(t)$ can survive due to the random phase mechanism.

The stationary phase approximation leads, 
\begin{eqnarray}\label{rpa}
|\Phi(t)\rangle\simeq\sum_{\nu_c}\tilde{\alpha}_{\nu_c}|\nu_c(t)\rangle e^{-i\Lambda_{\nu_c}(t)/\hbar}.
\end{eqnarray}
Here $\nu_c$ satisfies $\delta \Lambda_\nu(t)/\delta\nu =0.$

Since, in general, $\nu_c$ depends on time, the time dependence of 
$\Lambda_{\nu_c}(t)$ draws the envelope of $\{\Lambda_\nu(t)\}$.
Moreover, both the system $\phi$ and the environment $\varepsilon$ equivalently contribute to $\Lambda_\nu(t)$.
If we try to find out the classicality in our model, each of the branch $\nu_c$ should 
be considered as the independent reality.

\vspace{1cm}
\underline{\bf Example: localization of the center-of-mass}\\

For a short-time perturbation, 
$\Lambda_\nu(t)\simeq\langle\nu(t)|\hat{h}_I|\nu(t)\rangle\Delta t$ leads 
that the surviving state $|\nu_c(t)\rangle$ is the approximate eigenstate of the 
interaction Hamiltonian.
Especially 
when the states, $|\phi_1\rangle$ and $|\phi_2\rangle$, correspond to the position in space 
(namely $|\phi_i\rangle=|{\bf R}_i\rangle$), 
this means the localization of the COM. 
We here notice that the total state is still in the superposition of states. 

\vspace{1cm}
\underline{\bf Decoherence}\\

The eq. (\ref{rpa}) leads, 
\begin{eqnarray}
|\Phi(t)\rangle\simeq|\phi_1(t)\rangle\sum_{\nu^1_c}\tilde{\alpha}_{\nu^1_c}|\varepsilon_{\nu^1_c}(t)\rangle e^{-i\Lambda_{\nu^1_c}(t)/\hbar}+
|\phi_2(t)\rangle\sum_{\nu^2_c}\tilde{\alpha}_{\nu^2_c}|\varepsilon_{\nu^2_c}(t)\rangle e^{-i\Lambda_{\nu^2_c}(t)/\hbar},
\end{eqnarray}
where the state vectors, $|\phi_1(t)\rangle$ and $|\phi_2(t)\rangle$, have lost the coherence 
due to the factors, $\Lambda_{\nu^i_c}(t).$
If we can neglect the self-evolution of $\phi,$ 
the decoherence in Schmidt basis is also reproduced.
Especially, in the case that $\{\Lambda_\nu(t)\}$, as the function of $\nu$,  
has only two extreme values, $\Lambda_{\nu^1_c}(t)$ and $\Lambda_{\nu^2_c}(t)$, corresponding to $|\phi_1(t)\rangle$ and $|\phi_2(t)\rangle$, the above equation becomes, 
\begin{eqnarray}\label{rpa-two-states}
|\Phi(t)\rangle\simeq\tilde{\alpha}_{\nu^1_c}|\phi_1(t)\rangle|\varepsilon_{\nu^1_c}(t)\rangle e^{-i\Lambda_{\nu^1_c}(t)/\hbar}+
\tilde{\alpha}_{\nu^2_c}|\phi_2(t)\rangle|\varepsilon_{\nu^2_c}(t)\rangle e^{-i\Lambda_{\nu^2_c}(t)/\hbar}.
\end{eqnarray}
There are only two surviving states.

\vspace{1cm}
\underline{\bf Notice}\\

The above random-phase mechanism makes use of the entanglement between a system and its environment.
On the other hand, if we start from the product state, we obtain,  
\begin{eqnarray}
(c_1|\phi_1\rangle+c_2|\phi_2\rangle)(\alpha_1|\varepsilon_1\rangle+ \cdots+\alpha_N|\varepsilon_N\rangle)
=\sum_{\nu=1}^N\alpha_\nu(c_1|\phi_1\rangle+c_2|\phi_2\rangle)| \varepsilon_\nu\rangle.
\end{eqnarray}
Here each state  has same superposition state of the subsystem $\phi$
so that the random-phase mechanism does not work.
It, however, is natural that the interaction makes the coefficients $\{c_i\}$ depend on $\nu$ 
since the total system is repeatedly perturbed.

\begin{figure}
\begin{center}
\includegraphics[width=7cm]{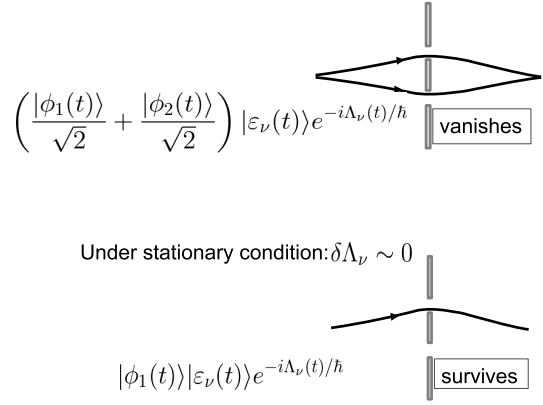}
\caption{The localized state corresponding to the stationary phase (image).}
\end{center}
\end{figure}

\newpage

\section{Conclusions}

We have studied the emergence of classical states under perturbative interaction. 
The states which interact with many other degrees of freedom, 
such as the COM of macro-objects, play important role.
Although the random phase mechanism is effective as same as Zurek's strong correlation model\cite{Zurek1982}, there are enormous states each of which independently developes due to 
the orthogonality of the environmental states.
In these privileged states, the subsystem picture with the separability is stable.

The time-integrated interaction energy, $\Lambda_\nu(t)$, 
comes into the phase of the states, the random phase mechanism selects out the states 
having extreme value of $\Lambda_\nu(t)$.
For short-time interaction, the eigenstate of the interaction Hamiltonian emerges.
In general, both the system $\phi$ and the environment $\varepsilon$ equivalently contribute to $\Lambda_\nu(t)$.

Since the random phase mechanism works also in our model,
for the privileged (pointer) states, the decoherence scenario can be reproduced.
Especially, in the case of eq. (\ref{rpa-two-states}),
the derivation of the decoherence  (in \S 2.3) seems accompanied with no coarse graining.
But examinations with practical models will be important.

\vspace{1cm}
\underline{\bf Beyond: classical complex systems}\\

A COM is the minimum unit forming the macroscopic order in our daily experience.
When two or more macro-objects are in single environment,
more complex classical system may emerge: 
\begin{eqnarray}
|\Phi(t)\rangle=\sum_{\nu_c}\tilde{\alpha}_{\nu_c}|\phi^A_{\nu_c}(t)\rangle|\phi^B_{\nu_c}(t)\rangle\cdots|\phi^Z_{\nu_c}(t)\rangle|\varepsilon_{\nu_c}(t)\rangle e^{-i\Lambda_{\nu_c}/\hbar}
\end{eqnarray}
If the direct interaction is absent among $A, B, \cdots, Z$, the time integral of the interaction energy is expressed as, $\Lambda_{\nu_c}=\Lambda^A_{\nu_c}+\Lambda^B_{\nu_c}+\cdots$.
It, however, is a future task to find the practical model for this problem.

\end{document}